\documentstyle[preprint,prl,aps,epsfig]{revtex}
\begin{document}
\draft
\narrowtext
\tighten
\title{Alternative evaluations of halos in nuclei}
\author{S. Karataglidis$^{(a)}$, P.~J.~Dortmans$^{(b)}$,
K.~Amos$^{(b)}$, and C.~Bennhold$^{(c)}$}
\address{%
$^{(a)}$ TRIUMF, 4004 Wesbrook Mall, Vancouver, British Columbia,
V6T 2A3, Canada\\
$^{(b)}$ School of Physics, University of Melbourne, Parkville,
Victoria, 3052, Australia\\
$^{(c)}$ Center of Nuclear Studies, Department of Physics, The George
Washington University, Washington, D.C., 20052}
\date{\today}
\maketitle
\begin{abstract}
Data for the scattering of $^6$He, $^8$He, $^9$Li, and $^{11}$Li from
hydrogen are analyzed within a fully microscopic folding model of
proton-nucleus scattering. Current data suggest that of these only
$^{11}$Li has a noticeable halo. For $^6$He, we have also analyzed the
complementary reaction $^6$Li($\gamma,\pi^+$)$^6$He$_{gs}$. The
available data for that reaction support the hypothesis that $^6$He
may not be a halo nucleus. However, those data are scarce and there is
clearly a need for more to elicit the microscopic structure of $^6$He.
\end{abstract}
\pacs{}

\section{Introduction}
Much information has been learned concerning the nature of halos in
nuclei from studies of heavy ion break up reactions in which the
momentum distributions of the valence nucleons have been found to be
very narrow \cite{Ha95}.  This observation suggests matter
distributions which extend well beyond the radius of the nuclear
potential and examples of halos found by this means are $^{11}$Li and
$^{11}$Be. Different neutron distributions in exotic nuclei, such as
skins ($^8$He, for example), also have been studied using this method.
However, doubt remains on the ability of such reactions to probe the
initial state wave functions. The breakup of $^6$He has been
demonstrated to be a two-step process \cite{Al98}, in which the $^5$He
fragment survives for a considerable amount of time as an $\alpha-n$
resonance before it breaks up. This suggests that the effects of final
state interactions are significant in this reaction, so that
information concerning the initial state wave function is lost.

Also, that approach has the disadvantage of missing part of the
initial state wave function of the halo nucleons \cite{Ha96} probing
only the asymptotic part of the wave function. Success has been
achieved in the analysis of those reactions using few-body models for
the halo nuclei (see Ref.~\cite{Ga99} and references therein) as they
are able to describe the asymptotic parts of nuclei better than most
shell models \cite{Ka97}. There remains the need to find ways of
studying microscopic properties of the wave functions of halo nuclei.

To study the microscopic aspects of the wave functions of exotic
nuclei we look to alternatives which probe the entire wave
function. Proton scattering in the inverse kinematics and charged pion
photoproduction are such reactions. Experiments have been performed
for the (elastic) scattering of radioactive ions from hydrogen (see,
for example, \cite{Ko97}). In the inverse kinematics this corresponds
to proton scattering from the heavy ion, which directly measures the
matter distribution of that ion. In particular, depending on the
momentum transfer, such scattering may measure the density near the
surface of the nucleus so that detailed information on the halo may be
collected.  Charged pion photoproduction from nuclei may serve as a
useful complementary probe of halo structures \cite{Ka98}, especially
as that reaction also is sensitive to the entire halo wave function
formed in the final state.  We present analyses of data on both of
these reactions to study the neutron distributions of $^6$He, $^8$He,
$^9$Li, and $^{11}$Li to determine whether the results permit
identification of any of these nuclei as a neutron halo or neutron
skin system.

\section{Models of structure}
As both proton scattering and charged pion photoproduction reactions
probe the microscopic structure of the nucleus, a suitable model for
the description of halo nuclear states in those reactions would be one
in which nucleon degrees of freedom are admitted. This would, by
necessity, include the core. In the case of $^{11}$Li scattering from
hydrogen, it was found that a full description of the $^9$Li core was
required \cite{Cr96} to describe the elastic scattering
data. Therefore, we describe the halo states within the shell model,
and allow for all nucleons to be active within the space (the
so-called ``no-core'' models).

Several groups report shell model calculations of $^{6,8}$He and
$^{9,11}$Li. Navr\'atil and Barrett \cite{Na96,Na98} have made
large-space shell model calculations using interactions obtained
directly from the $NN$ $g$ matrices, with the Reid93 $NN$ interaction
as their base. Their calculations for $^6$He were made in a complete
$(0+2+4+6)\hbar\omega$ model space while those for $^8$He, $^9$Li and
$^{11}$Li were made in the smaller $(0+2+4)\hbar\omega$ model space;
the limitation arising from the dimensionality increasing with mass
for a given space. (Henceforth, only the highest excitation will be
given in reference to the complete model space.) Good results were
found for the ground state properties in each case. For $^6$He,
specifically, their calculations indicate that there is little or no
need for this system to have a neutron halo to obtain agreement. For
the other nuclei, they find spectra and ground state properties that
are also quite good, although the calculated proton root-mean-square
(r.m.s.) radii are small in comparison to the measured values. The cause
of these discrepancies may be a halo-like distribution of the excess
neutrons; the $4\hbar\omega$ model space is not large enough to admit
such halo characteristics for these nuclei \cite{Na98}. These
calculations may be contrasted with the results of our recent study
\cite{Ka97a} in which the results of $0\hbar\omega$ and $2\hbar\omega$
shell model calculations of $^9$Li and $^{11}$Li, made using
phenomenological interactions, were reported. When using the wave
functions obtained in those smaller space calculations, the available
elastic scattering data at $60A$ and $68A$~MeV from hydrogen were well
described.

We have calculated the wave functions for $^{6,8}$He within a complete
$4\hbar\omega$ model space using the $G$ matrix interaction of Zheng
{\em et al.} \cite{Zh95}. For $^{9,11}$Li, we used the wave functions
as calculated in our previous work \cite{Ka97a}: using the P$(5-16)$T
interaction in the $0\hbar\omega$ model space for $^9$Li, and the WBP
interaction \cite{Wa92} in the $2\hbar\omega$ model space for
$^{11}$Li. All calculations were made using the shell model code
OXBASH \cite{Ox86}. From those wave functions, the one-body density
matrix elements (OBDME) were obtained to use in the descriptions of
the scattering and of the ($\gamma,\pi^+$) reaction.

\begin{figure}
\centering\epsfig{file=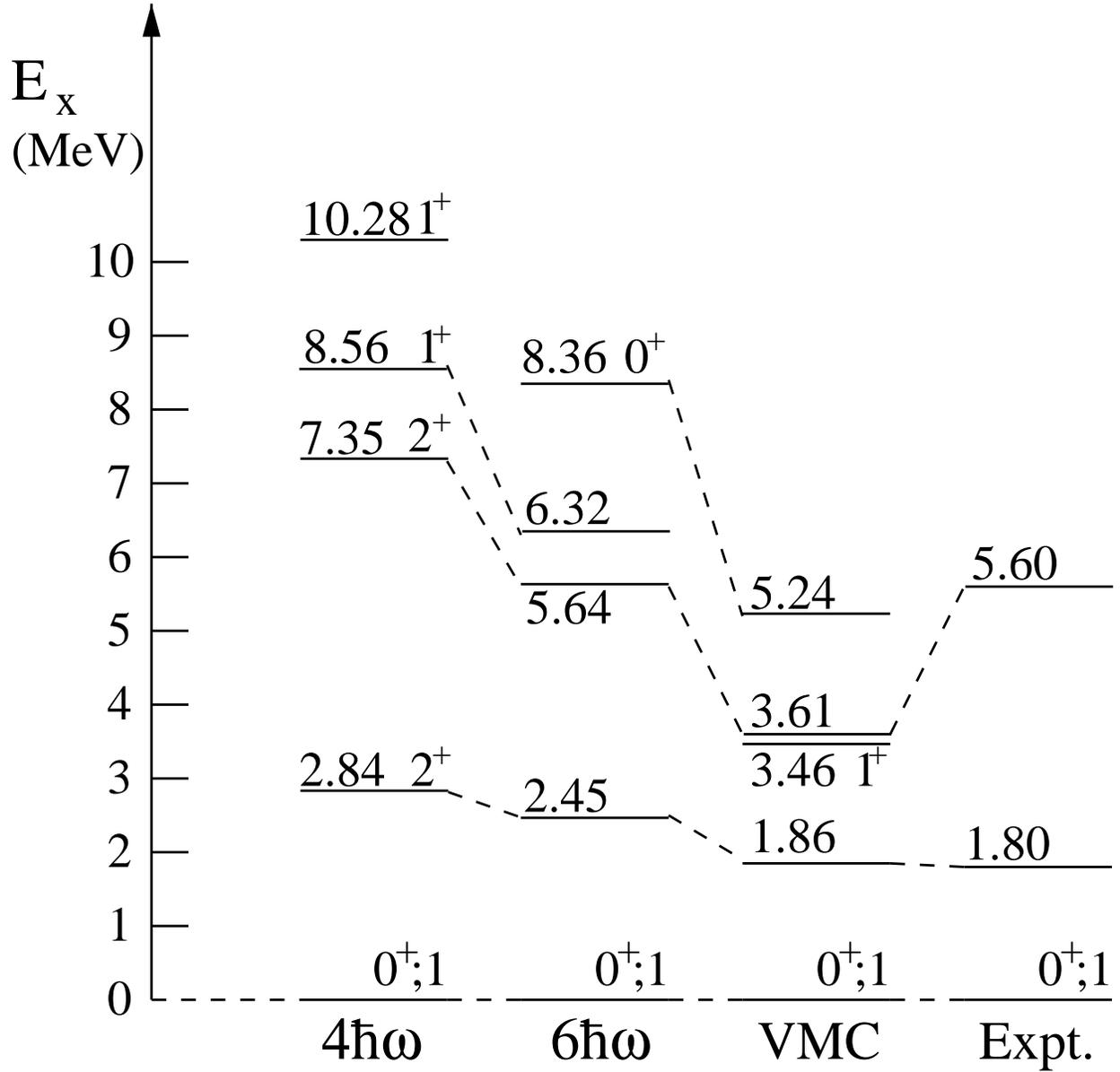,width=\linewidth,clip=}
\caption[]{The spectrum of $^6$He. The result of our $4\hbar\omega$
shell model calculation is compared to that of the $6\hbar\omega$
calculation \cite{Na96}, that of the VMC calculation \cite{Pu97}, and
to experiment \cite{Ja96}.}
\label{he6spec}
\end{figure}

The spectrum of $^6$He is displayed in Fig.~\ref{he6spec}. Therein,
the results of our calculation are compared to those of the
$6\hbar\omega$ calculation of Navr\'atil and Barrett \cite{Na96}, as
well as to those of Pudliner {\em et al.}  \cite{Pu97}, in which the
spectra of $A = 6$ nuclei were calculated using the Variational Monte
Carlo (VMC) shell model approach. The experimental spectrum was
obtained from Ref. \cite{Ja96}. The two calculations made using the
``traditional'' shell model approach ascribe $J^{\pi};T = 2^+;1$ to
the first two excited states, in agreement with experiment. While the
energy of the $2^+_1;1$ state is similar in the $4\hbar\omega$ and
$6\hbar\omega$ models, the energy of the $2^+_2$ state in the
$6\hbar\omega$ model is in much better agreement with the data. This
may be due to the modification of the auxiliary potential in the
Hamiltonian in that calculation \cite{Na96}. Without that
modification, overbinding is observed, of the order of 4~MeV. However,
it does not affect the spectrum significantly; the increase in energy
of each state is less than 1~MeV. It should be noted that this
overbinding will also affect our calculations, as we use the same
interactions, although we do not expect that the wave functions will
be significantly affected. The results of the VMC calculation place
the $2^+_1$ state in very close agreement with experiment. However,
that calculation also has an extra $1^+$ state in the spectrum not
observed, nor seen in the other calculations.  It would be interesting
to investigate in more detail the character of that particular state.

\begin{figure}
\centering\epsfig{file=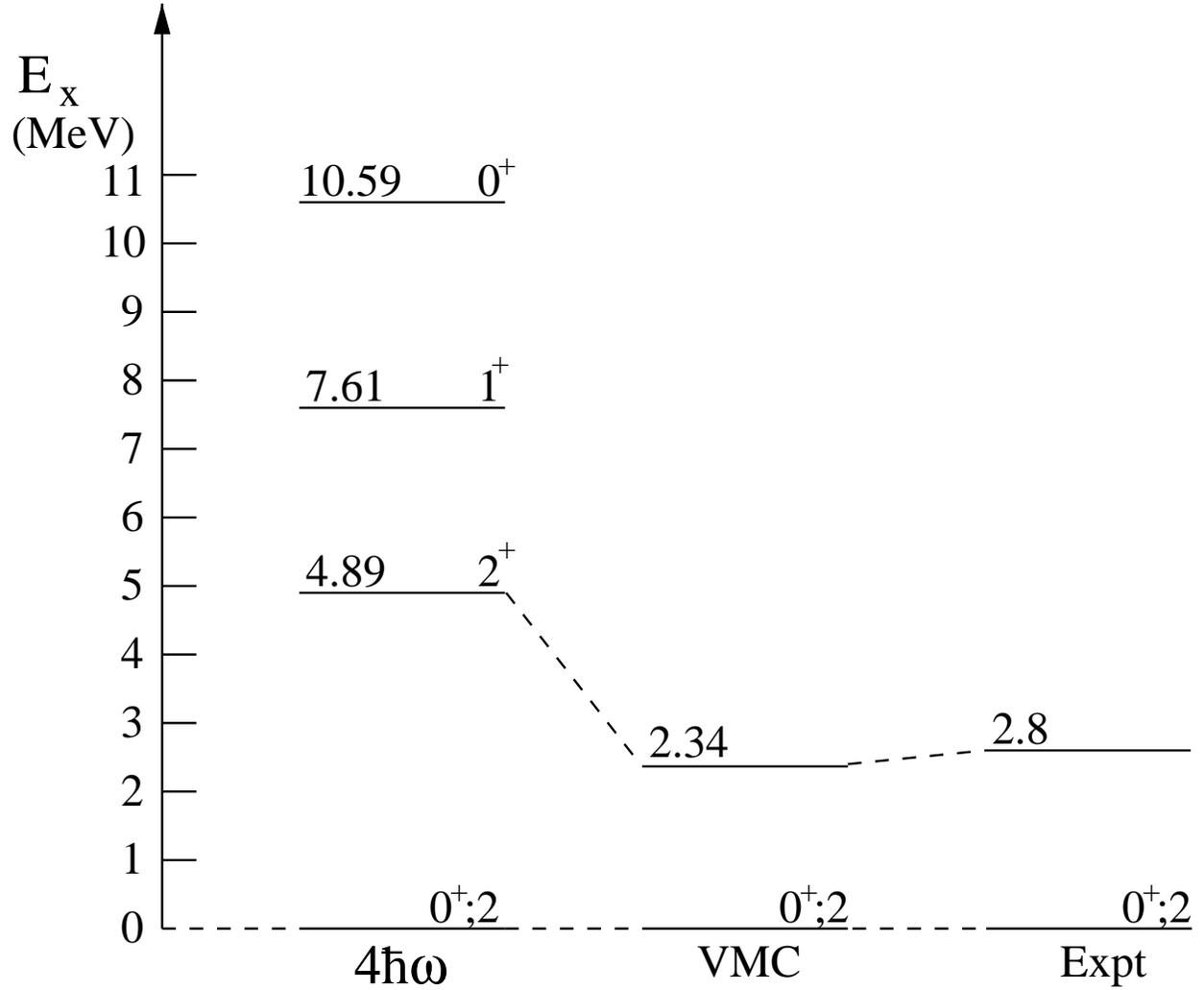,width=\linewidth,clip=}
\caption[]{The spectrum of $^8$He. The result of the present
$4\hbar\omega$ shell model calculation is compared to that of the VMC
calculation \cite{Wi98}. The data are from Ref.~\cite{Aj88}.}
\label{he8spec}
\end{figure}

There is very little experimental information on the spectrum of
$^8$He. The first excited state is listed at $2.8 \pm 0.4$~MeV and has
$J^{\pi};T = (2^+);2$ \cite{Aj88}. Other states are reported at 1.3, 2.6
and 4.0~MeV \cite{Aj88}, as obtained from a transfer experiment
involving heavy ions, but no other data are available as yet to
support those measurements. The results from the present calculation
are compared to those obtained from the VMC calculation \cite{Wi98} in
Fig.~\ref{he8spec}. The spectrum obtained by Navr\'atil and Barrett in
the $4\hbar\omega$ model space using their updated $G$ matrix
interaction \cite{Na98} is similar to the present results, and so are
not shown. The $2^+_1;2$ state is predicted correctly by all
calculations as the first excited state, although only the VMC
calculation agrees well with experiment. The disagreement between the
shell model calculations and experiment may be due to the shell model
failing to reproduce, within the $4\hbar\omega$ model space, the
correct neutron density distribution. $^8$He has a well-known neutron
skin, the description of which may require a calculation using a very
large model space.

\begin{figure}
\centering\epsfig{file=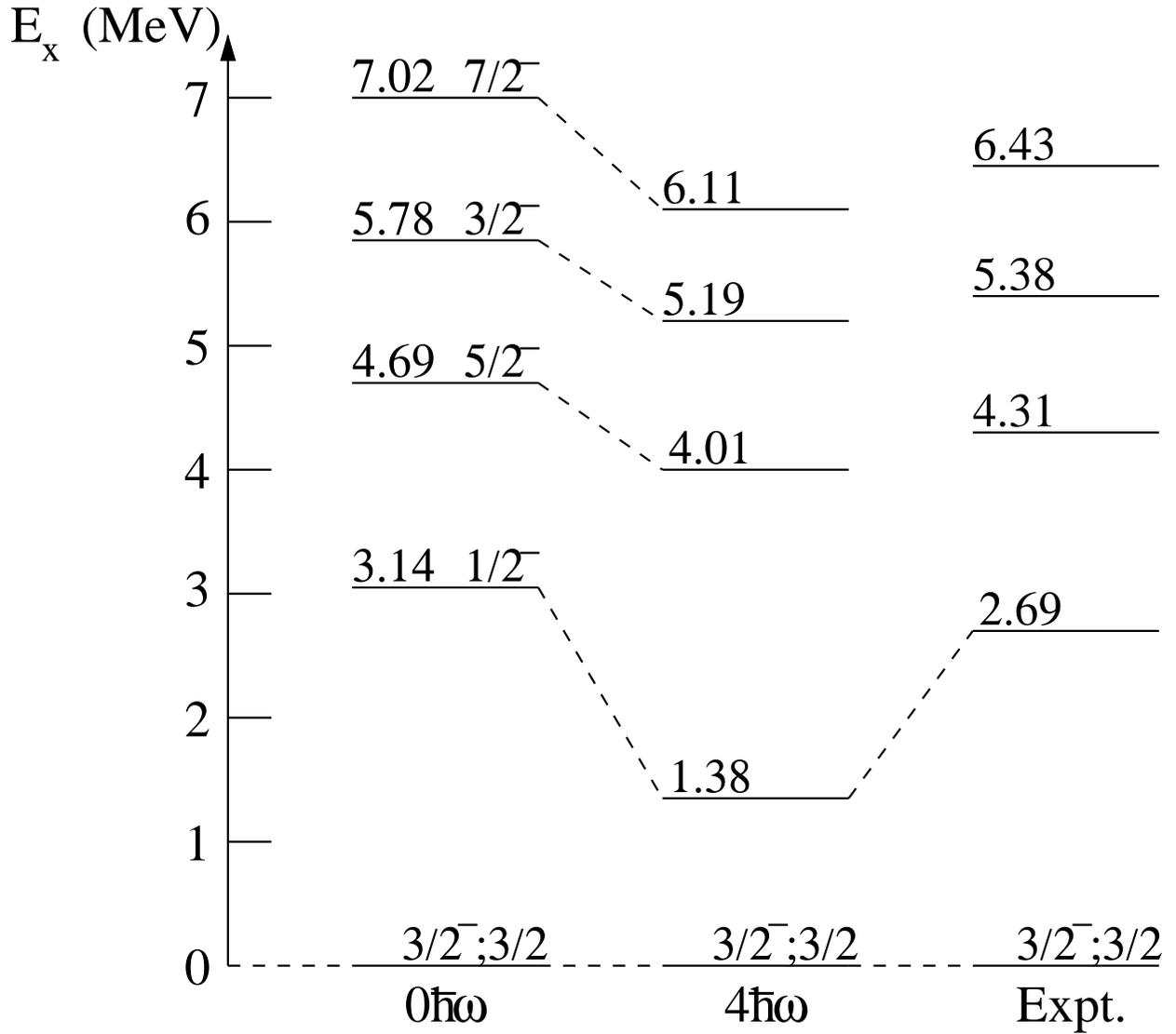,width=\linewidth,clip=}
\caption[]{The spectrum of $^9$Li. The result of the present
$0\hbar\omega$ shell model calculation is compared to that obtained in
the $4\hbar\omega$ model space \cite{Na98}. The data are from
Ref.~\cite{Aj88}.}
\label{li9spec}
\end{figure}

The $^9$Li spectrum is displayed in Fig.~\ref{li9spec}, wherein the
results of the present calculation are compared to those obtained
within the $4\hbar\omega$ model space. The experimental energies are
obtained from \cite{Aj88}. The spectrum obtained in the $0\hbar\omega$
model space is in general agreement with that obtained in the
$4\hbar\omega$ model space, although the first excited state comes
much lower in the latter. There are no spins assignments in the
experimental spectrum bar the ground and first excited states, which
the models correctly predict. As we consider only the elastic channel
in the calculations of proton scattering, the $0\hbar\omega$
calculation is sufficient. One expects that core polarization
corrections will become important for inelastic scattering.

\begin{figure}
\centering\epsfig{file=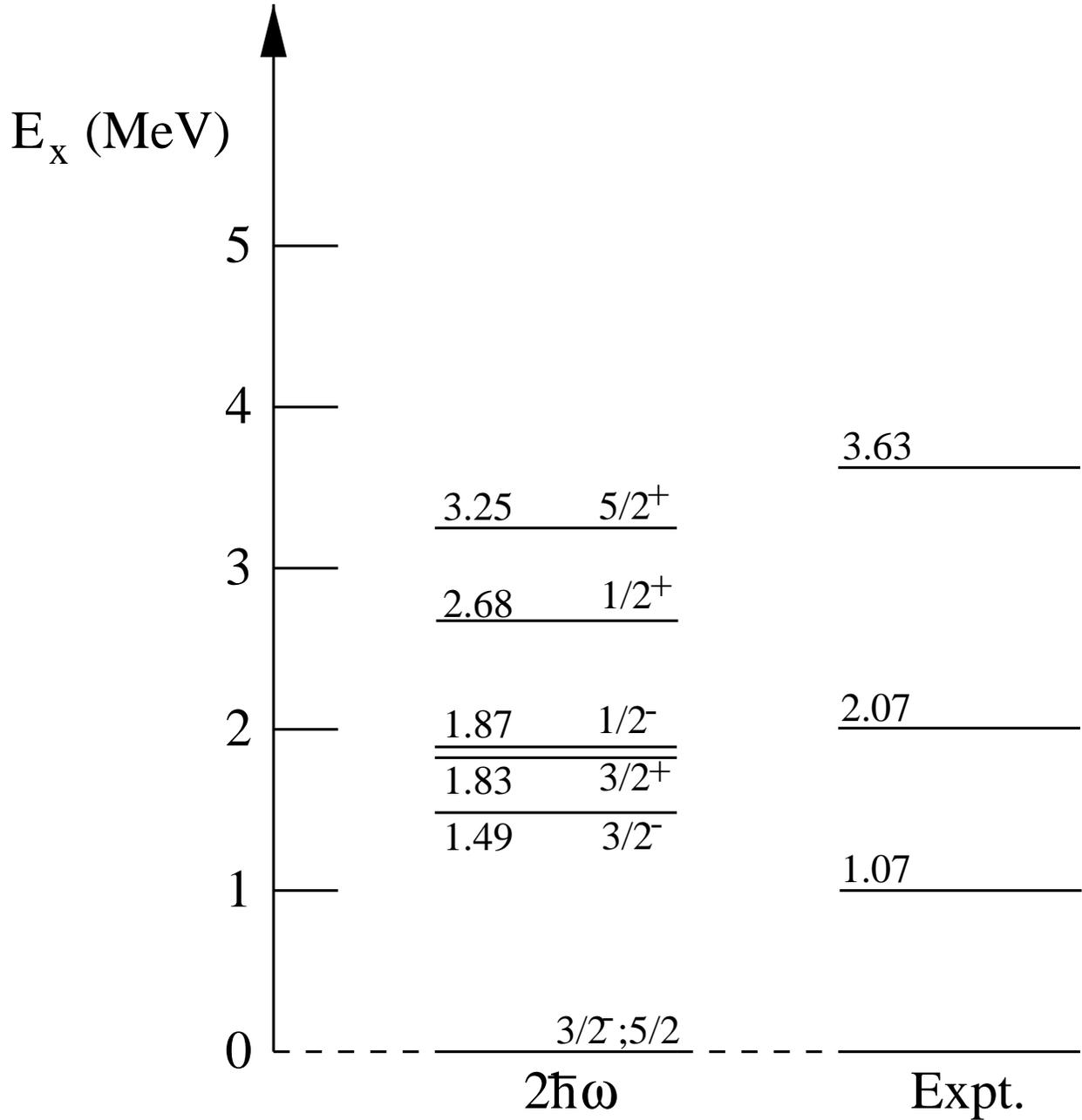,width=\linewidth,clip=}
\caption[]{The spectrum of $^{11}$Li. The result of the present
$2\hbar\omega$ shell model calculation is compared to the data of
Gornov {\em et al.} \cite{Go98}.}
\label{li11spec}
\end{figure}

The $^{11}$Li spectrum is displayed in Fig.~\ref{li11spec}. Therein,
the experimental results of Gornov {\em et al.} \cite{Go98} are
compared to the results of the present calculation. The experiment
from which the excitation spectrum was obtained was
$^{14}$C($\pi^-,pd$)$^{11}$Li and did not allow for any spin
assignments to be made so the comparison between experiment and theory
at this stage must be tentative. The $\frac{1}{2}^-_1;\frac{5}{2}$
state is formed from the coupling of the valence neutrons to the
$\frac{1}{2}^-$ state in $^9$Li.

\section{Elastic proton scattering}
We now consider elastic scattering of the heavy ions from hydrogen,
data for which are available at $72A$~MeV for $^{6,8}$He and $62A$~MeV
for $^{9,11}$Li. The analyses follow those made for the elastic
scattering of 65~MeV protons from various targets ranging from $^6$Li to
$^{238}$U \cite{Do98}, and we refer the reader to that reference for
complete details. We present a brief summary of the formalism herein.

There are three essential ingredients one must specify to calculate
proton scattering observables. The first are the OBDME as obtained
from the shell model calculations. They are explicitly defined as
\begin{equation}
S_{\alpha_1 \alpha_2 I} = \left\langle J_f \left\| \left[
a^{\dagger}_{\alpha_2} \times \tilde{a}_{\alpha_1} \right]^I \right\|
J_i \right\rangle
\end{equation}
where $J_i$ and $J_f$ are the initial and final nuclear states
respectively, $I$ is the angular momentum transfer, and $\alpha_i =
\left\{ n_i, l_i, j_i, \rho_i \right\}$ with $\rho$ specifying either a
proton or a neutron.

The second ingredient is the effective interaction between the
projectile nucleon and each and every nucleon in the target. The
complex, fully nonlocal, effective interaction we choose \cite{Do98}
accurately maps onto a set of nucleon-nucleon ($NN$) $g$
matrices. These density-dependent $g$ matrices are solutions of the
Brueckner-Bethe-Goldstone equations in which a realistic $NN$
potential defines the basic pairwise interaction. For that, we have
chosen the Paris interaction \cite{La80}. Good to excellent
predictions of the elastic scattering observables for stable targets
from $^6$Li to $^{238}$U were found with this effective (coordinate
space) interaction.

Finally, the single particle wave functions describing the nucleon
bound states must be specified. For the present calculations we
distinguish between those calculations which yield an extensive (halo)
density distribution and those that do not. The former we designate
``halo'' while the latter are designated ``non-halo''.  Those
calculations use single-particle wave functions as specified naively
from the shell model calculations, which do not make allowance
directly for the very loose binding of the valence neutrons, at least
not to the level in $\hbar\omega$ assumed in the model spaces. In all
cases bar one, Woods-Saxon (WS) wave functions were used. Those which
gave good reproduction of the elastic electron scattering form factors
of $^6$Li \cite{Ka97} were used for all the $^{6,8}$He calculations
while those which reproduced the elastic electron scattering form
factors of $^9$Be \cite{Do97} were used in the calculations for
$^9$Li, and also for the core in the halo calculation of
$^{11}$Li. For the non-halo specification of $^{11}$Li, we used
appropriate harmonic oscillator wave functions for mass-11
\cite{Ka97a}. To specify the halo, we adjusted the WS potentials from
the values given such that the relevant valence neutron orbits are
weakly bound. Those are the $0p$-shell orbits and higher for the
helium isotopes, and the $0p_{\frac{1}{2}}$ orbit and higher for the
lithium ones. Such an adjustment to single particle wave functions
adequately explains the very large $B(E1)$ in $^{11}$Be \cite{Mi83}
and guarantees an extensive neutron distribution. In our analyses,
$^8$He and $^9$Li act as controls: $^8$He is an example of a neutron
skin and $^9$Li is a simple core nucleus. The single neutron
separation energies are 2.137~MeV and 4.063~MeV for $^8$He \cite{Aj88}
and $^9$Li \cite{Aj90}, respectively. We may artificially ascribe a
halo to these nuclei, by setting a much lower separation energy, to
ascertain if the procedure and data are sensitive enough to detect the
flaw. For $^6$He, the 0p-shell binding energy was set to 2~MeV, which
is close to the separation energy (1.87~MeV \cite{Aj88}) of a single
neutron from $^6$He, leaving the lowest $0p$-shell resonance in
$^5$He. For $^8$He, $^9$Li and $^{11}$Li, the halo was specified by
setting the binding energy for the WS functions of the
$0p_{\frac{1}{2}}$ and higher orbits to 0.5~MeV \cite{Ka97a}.  While
the halo and nonhalo specifications are a matter of convenience at
this point, we test the validity of halo name by calculating the r.m.s.
radii for all four nuclei.

The ability by which the wave functions can describe halo states may
be evaluated by calculating the r.m.s. radius for each nucleus and
compare to those results obtained from analyses of the reaction cross
sections. The r.m.s. radii are presented in Table~\ref{radii}, as
calculated using the shell model wave functions and the specified
single particle wave functions. The values obtained from the shell
model using the correct single particle wave functions are largely
consistent with those obtained from few-body calculations
\cite{To97,Al96,Al98a}. The values obtained indicate that $^6$He and
$^{11}$Li are halo nuclei, while $^8$He and $^9$Li are not. While our
prediction for the r.m.s. radius for $^{11}$Li appears lower compared
to the value extracted from the reaction cross section \cite{Al96}, it
is consistent within the error bars quoted with that analysis. The
lower value may be due to the wave functions possibly being incapable
of describing long range phenomena adequately. If that is the case,
more $\hbar\omega$ excitations must be admitted into the model space;
although the present set of wave functions should be sufficient to
describe the proton scattering observables at high momentum transfer.

The calculations for the scattering from $^9$Li and
$^{11}$Li are those presented in Ref.~\cite{Ka97a}, while those for
$^6$He and $^8$He used the OBDME as we have obtained from our shell
model wave functions.

\begin{figure}
\centering\epsfig{file=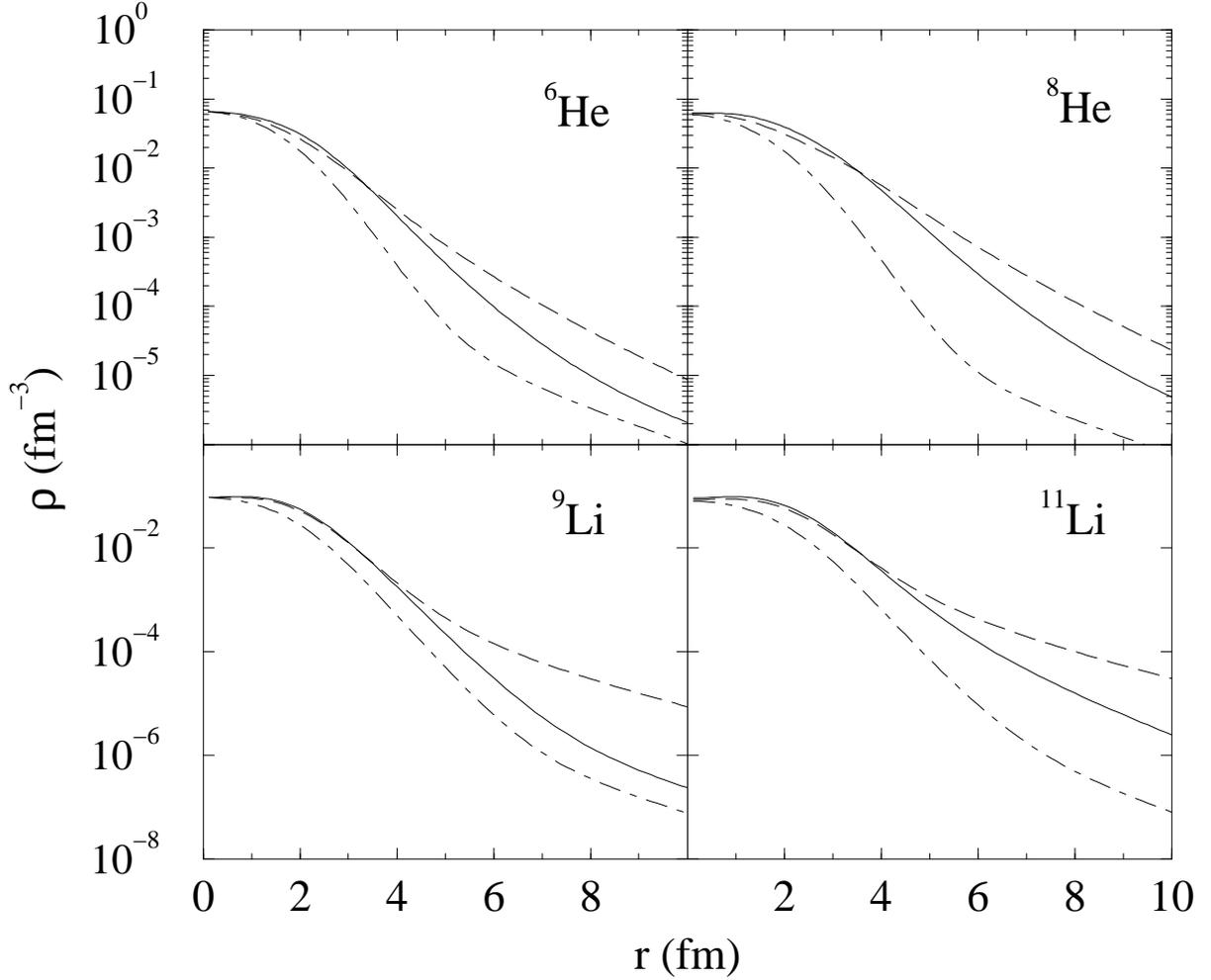,width=\linewidth,clip=}
\caption[]{The (shell model) neutron density profiles for the nuclei
$^{6,8}$He and $^{9,11}$Li. The dashed and solid curves represent,
respectively, the profiles when a halo is and is not contained in
those structures. The dot-dashed lines represent the proton density
for each nucleus.}
\label{fig1}
\end{figure}

The neutron density profiles for $^6$He, $^8$He, $^9$Li, and $^{11}$Li
obtained from the present shell model calculations are shown in
Fig.~\ref{fig1}.  Therein the dashed and solid lines portray,
respectively, the profiles found with and without the halo conditions
being implemented. The dot-dashed line in each case represents the
proton density. As the folding process defines the optical potentials,
the internal ($r < r_{\rm rms}$) region influences the predictions of
differential cross sections, notably at large scattering angles. In
this region the extensive (halo) distribution exhibits a lower density, as
the neutron strength is bled to higher radii. That effect
characterized the proton halo in $^{17}$F$^{\ast}$ as manifest in the
$^{17}$O($\gamma,\pi^-$) reaction \cite{Ka98}. The extended nature of
the halo also influences the optical potentials as evidenced in
changes to the cross sections at small momentum transfers (typically
$< 0.5$~fm$^{-1}$ or $\theta_{c.m.} < 15^{\circ}$ for beam energies
between $60A$ and $70A$~MeV).

\begin{figure}
\centering\epsfig{file=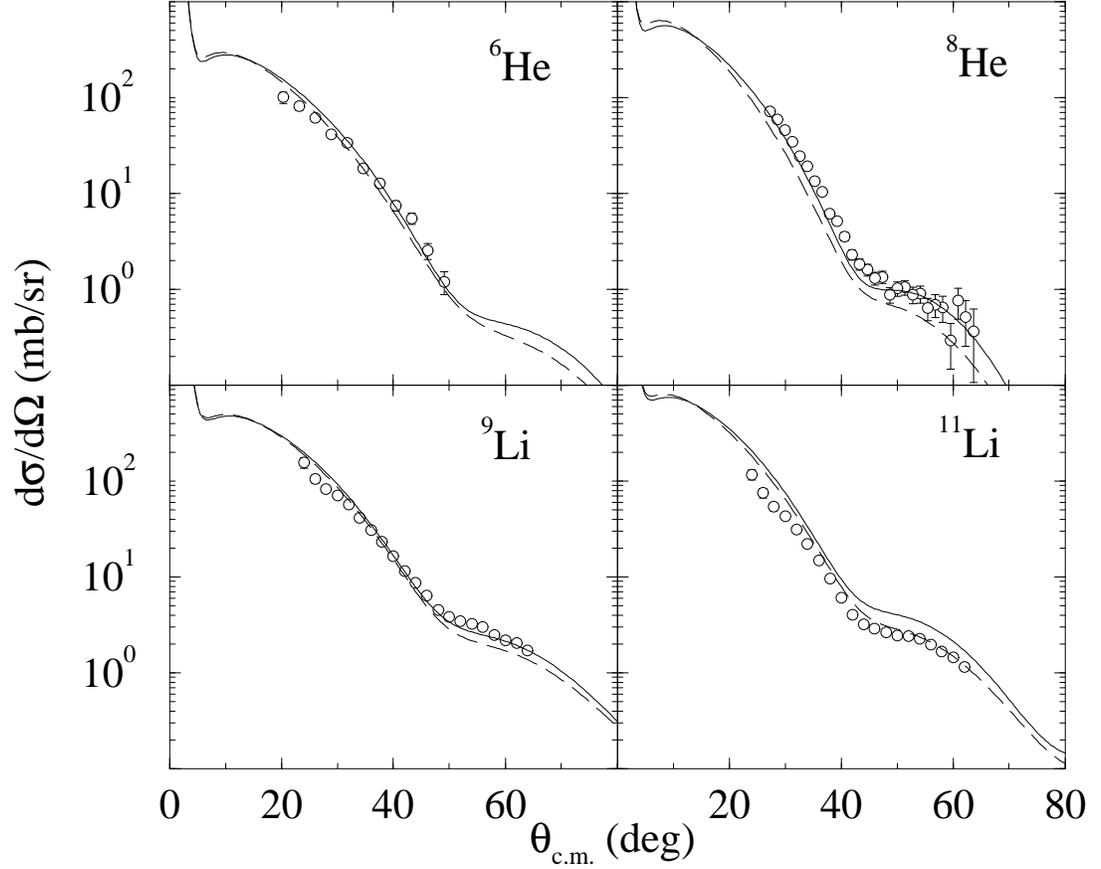,width=\linewidth,clip=}
\caption[]{Predictions of the differential cross sections from the
scattering of 72$A$~MeV $^{6,8}$He and of 62$A$~MeV $^{9,11}$Li from
hydrogen compared with experimental data. The data are from
Refs. \cite{Ko93,Ko97,Mo92} and the results, assuming that each
nucleus has (does not have) a halo structure, are portrayed by the
dashed (solid) curves.}
\label{fig2}
\end{figure}

\begin{figure}
\centering\epsfig{file=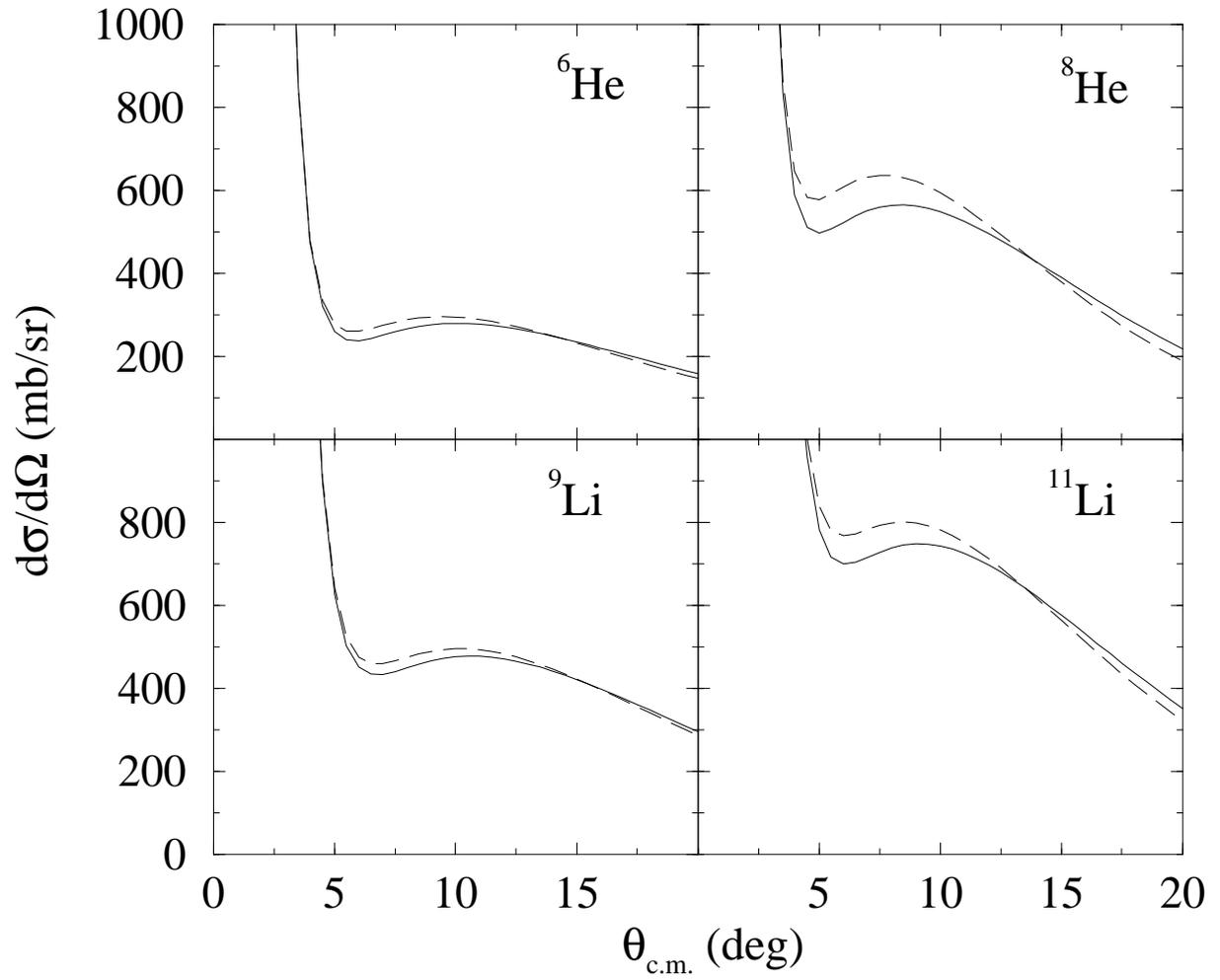,width=\linewidth,clip=}
\caption[]{Differential cross sections as shown in Fig.~\ref{fig2},
but for small angles only.}
\label{fig3}
\end{figure}

The predicted differential cross sections for the scattering of
$^{6,8}$He and $^{9,11}$Li from hydrogen are presented in
Figs.~\ref{fig2} and \ref{fig3}. In Fig.~\ref{fig2} we display the
results to $80^{\circ}$ ($q \sim 2.5$~fm$^{-1}$) and compare them with
the data taken by Korsheninnikov {\em et al.} \cite{Ko97,Ko93} using
$70.5A$~MeV $^6$He and $72A$~MeV $^8$He beams, and by Moon {\em et
al.} \cite{Mo92} using $60A$~MeV $^9$Li and $62A$~$^{11}$Li beams. The
forward angle results specifically, for which there are no data, are
shown in Fig.~\ref{fig3} to emphasize the influence on the predictions
by the extension of the halo ($r > r_{\rm rms}$). In both figures the
solid curves depict the non-halo results while the dashed curves are
those with the halo.

As is evident in Fig.~\ref{fig2}, the data for our two controls,
$^8$He and $^9$Li, are sufficient to resolve the question of whether
these nuclei exhibit halos. In both cases the data above $50^{\circ}$
are reproduced by the non-halo results suggesting that these nuclei do
not have extended (halo) neutron distributions. This gives confidence in our
ability to use such data to determine if a nucleus has a halo. That is
confirmed in the case of the scattering of $^{11}$Li from hydrogen as
the data clearly support a halo structure. There are differences
evident between the halo and the non--halo predictions with these
nuclei when one considers small angle scattering, where the influence
of the Coulomb interaction is quite important. We present the results
of our calculations for small angle scattering in Fig.~\ref{fig3}.
For $^9$Li, the difference between the halo and non-halo results is
small which supports the notion that this nucleus is a close-packed
system. This is contrasted by the results for both $^8$He and
$^{11}$Li: the difference between the halo and non-halo results for
$^{11}$Li is greater, suggesting again the halo structure, but the
difference is greatest in $^8$He. Together with the large angle
scattering data this suggests the neutron skin structure for $^8$He
serves to dilute the charge distribution stemming from the two protons
while pushing the density of the neutrons uniformly to larger radii as
is shown in Fig.~\ref{fig1}.

We now turn our attention to $^6$He. As shown in Fig.~\ref{fig2}, the
existing $^6$He data range only to $50^{\circ}$ ($q \sim
1.6$~fm$^{-1}$). This is insufficient to discriminate between the halo
and non-halo structures. As confirmed by the data and optical model
analysis of Korsheninnikov {\em et al.} \cite{Ko97}, our results are
almost identical to those from $p$--$^6$Li scattering, but only in the
region where the data were taken for the $p$--$^6$He
scattering. Beyond this region there is a sufficient difference
between the calculations to determine if $^6$He exhibits a halo. Data
are needed beyond $50^{\circ}$ to make such an assessment. The small
angle scattering shown in Fig.~\ref{fig3} is consistent with the
result for $^9$Li in showing little difference between the halo and
non-halo results.

\begin{figure}
\centering\epsfig{file=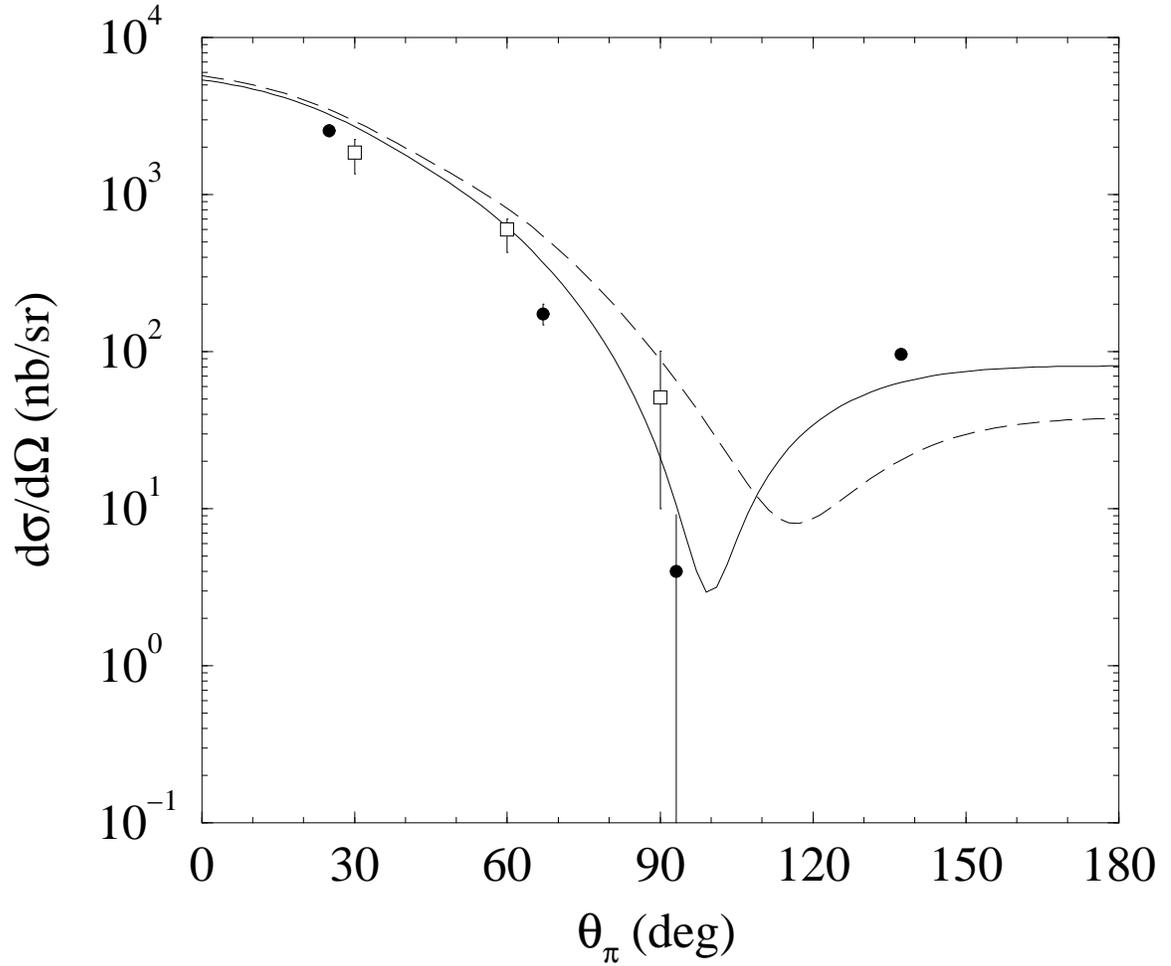,width=\linewidth,clip=}
\caption[]{The $^6$Li($\gamma,\pi^+$)$^6$He reaction for $E_{\gamma} =
200$~MeV. The data of Shaw {\em et al.} (circles) \cite{Sh91} and
Shoda {\em et al.} \cite{Sh81} (squares) are compared to the results
with and without halo as displayed by the dashed and solid lines.}
\label{fig4}
\end{figure}

We may also study $^6$He via the $^6$Li($\gamma,\pi^+$)$^6$He$_{gs}$
reaction. This reaction may be more sensitive to details of the halo
as the transition is more sensitive to the descriptions of the valence
neutrons. We have calculated the cross sections for this reaction at
$E_\gamma = 200$~MeV using the DWIA model of Tiator and Wright
\cite{Ti84}. As the $^6$He ground state is the isobaric analogue of
the $0^+;1$ (3.563~MeV) state in $^6$Li, we have used the OBDME for
the transition to that state in $^6$Li, as obtained from a complete
$(0+2+4)\hbar\omega$ shell model calculation \cite{Ka97}. The non-halo
result corresponds to a calculation using harmonic oscillator
single-particle wave functions with $\hbar\omega = 12.65$~MeV
\cite{Ka97}. Those wave functions are also used for the initial $^6$Li
state to obtain the halo result with the final $^6$He state being
specified by WS wave functions in the $0p$-shell and higher orbitals
only as given in the halo calculation of the scattering presented
above. Such a specification introduces a problem in normalization with
the $0p_{\frac{3}{2}}$ wave functions. The overlap of the HO and WS
$0p_{\frac{3}{2}}$ radial wave functions is 0.96, hence the wave
functions preserve the norm to within 4\%. Both results are compared
to the data of Shaw {\em et al.}  \cite{Sh91} (circles) and Shoda {\em
et al.}  \cite{Sh81} (squares) in Fig.~\ref{fig4}, wherein the halo
and non-halo calculations are displayed by the dashed and solid lines
respectively. From the available data one may infer that the non-halo
result is favored, but this is due to the datum at $137^{\circ}$
only. Note that our non-halo result is similar to that found by Doyle
{\em et al.} \cite{Do95} in which they used a $0\hbar\omega$ model of
structure and no specific halo structure was specified. Our halo
result is very similar to the result obtained from a three-body
description of $^6$He in which the wave functions reproduced the halo
properties \cite{Er99}. More data in
the region of the possible minimum as well as at large angles are
needed to confirm the conjecture that $^6$He does not have a halo
structure.

\section{Conclusions}
The available scattering data from hydrogen confirm that
$^{11}$Li is a halo nucleus, while the analysis of the scattering data
correctly determines that both $^8$He and $^9$Li are not. This
confirms our ability to predict correctly any halo structures as
probed by the scattering of exotic nuclei from hydrogen. The
low-angle scattering results also suggest that $^8$He is a neutron
skin nucleus, as found from breakup reactions.

While the data on the r.m.s. radii suggests that $^6$He is a halo
nucleus, the available scattering data for $^6$He from hydrogen are
not extensive enough to discriminate between the halo and non-halo
scenarios; in the measured region they suggest for $^6$He a very
similar matter distribution compared to $^6$Li. The complementary
$^6$Li($\gamma,\pi^+$)$^6$He reaction data suggest the non-halo
hypothesis. However, it must be stressed that more data, particularly
involving transitions to states in $^6$He, are required to support or
refute this conjecture.

The analysis presented here also demonstrates that, to test structure
models of these exotic nuclei most intensively, one should study
reactions of skin and halo nuclei with complementary probes and in
complementary reaction channels.

Financial support from the Natural Sciences and Engineering Research
Council of Canada, the Australian Research Council, and Department of
Energy Grant no. DE-FG02-95ER-40907 is gratefully acknowledged.

%
%
\begin{table}
\caption[]{Root-mean-square (r.m.s.) radii in fm for $^6$He, $^8$He,
$^9$Li, and $^{11}$Li. The results of our shell model calculations are
compared to those obtained from a Glauber model analysis of the
reaction cross sections \cite{Al96,To97}, and also from a few-body model
analysis of scattering data from hydrogen \cite{Al98}.}
\label{radii}
\begin{center}
\begin{tabular}{crrr}
Nucleus & \multicolumn{3}{c}{$r_{r.m.s.}$} \\
        & non-halo & halo & Glauber model \\
\hline
$^6$He & 2.301 & 2.586 & $2.54 \pm 0.04$ \\
$^8$He & 2.627 & 2.946 & 2.60$^{(a)}$ \\
$^9$Li & 2.238 & 2.579 & $2.30 \pm 0.02$ \\
$^{11}$Li & 2.447 & 2.964 & $3.53 \pm 0.10$ \\
\end{tabular}
\footnotesize{(a) Taken from \cite{Al98}.}
\end{center}
\end{table}

%
%

\end{document}